\documentclass[%
reprint,
superscriptaddress,
showpacs,preprintnumbers,
amsmath,amssymb,
aps,
pra,
longbibliography
]{revtex4-2}

\usepackage{amsmath,amssymb}
\usepackage{graphicx}
\usepackage{bm}
\usepackage{mathrsfs}
\usepackage{dsfont}

\usepackage[colorlinks,allcolors=blue]{hyperref}

\makeatletter

\newcommand{\Rmnum}[1]{\expandafter\@slowromancap\romannumeral #1@}
\newcommand{\mI}{\mathcal{I}}
\newcommand{\mJ}{\mathcal{J}}
\newcommand{\mB}{\mathcal{B}}
\newcommand{\mC}{\mathcal{C}}
\newcommand{\mM}{\mathcal{M}}
\makeatother

\begin{document}
\title{Verification of Kochen-Specker-type quantum contextuality with a single photon}

\author{Shihao Ru}
\email{shihao.ru.2018@gmail.com}
\affiliation{Shaanxi Key Laboratory of Quantum Information and Quantum Optoelectronic Devices, School of Physics, Xi'an Jiaotong University, Xi'an 710049, China}
\author{Weidong Tang}
\affiliation{School of Mathematics and Statistics, Shaanxi Normal University, Xi'an 710119, China}
\author{Yunlong Wang}
\affiliation{Shaanxi Key Laboratory of Quantum Information and Quantum Optoelectronic Devices, School of Physics, Xi'an Jiaotong University, Xi'an 710049, China}
\author{Feiran Wang}
\affiliation{Shaanxi Key Laboratory of Quantum Information and Quantum Optoelectronic Devices, School of Physics, Xi'an Jiaotong University, Xi'an 710049, China}
\affiliation{School of Science, Xi'an Polytechnic University, Xi'an 710048, China}
\author{Pei Zhang}
\affiliation{Shaanxi Key Laboratory of Quantum Information and Quantum Optoelectronic Devices, School of Physics, Xi'an Jiaotong University, Xi'an 710049, China}
\author{Fuli li}
\email{flli@mail.xjtu.edu.cn}
\affiliation{Shaanxi Key Laboratory of Quantum Information and Quantum Optoelectronic Devices, School of Physics, Xi'an Jiaotong University, Xi'an 710049, China}
\date{\today}

\begin{abstract}
Contextuality provides one of the fundamental characterizations of quantum phenomena, and can be used as a resource in lots of quantum information processing.
In this paper, we summarize and derive some equivalent noncontextual inequalities from different noncontextual models of the proofs for Kochen-Specker theorem based on Greenberger-Horne-Zeilinger states.
These noncontextual inequalities are equivalent up to some correlation items which hold both for noncontextual hidden variable theories and quantum mechanics.
Therefore, using single-photon hyperentangled Greenberger-Horne-Zeilinger states encoded by spin, path and orbital angular momentum, we experimentally verify several state-dependent noncontextual models of the proofs for the Kochen-Specker theorem by testing an extreme simplest Mermin-like inequality.
\end{abstract}

\maketitle
\section{Introduction}
Nonlocality and contextuality are important properties of quantum mechanics (QM) and are essential points different from classical mechanics. There are some efforts which try to put these characters into a deterministic form by including hidden variables.
Two typical hidden variable theories, local hidden variable theories (LHVTs) and noncontextual hidden variable theories (NCHVTs), have been proposed \cite{Bell64,KS67,CHSH,GHZ89}. 
LHVTs assume that all natural processes are local, information and correlations are propagated at most at the speed of light. The hidden aspect ascribes variability of experimental results to uncontrollable parameters of the model.
NCHVTs state that the predefined value of an observable is independent of any contexts (a set of compatible observables) which it is simultaneously measured with. Proofs of these hidden variable theories can be summed up in two famous theorems, i.e. Bell’s theorem \cite{Bell64} and Kochen-Specker (KS) theorem \cite{KS67}.
Bell's theorem states that no LHVTs can reproduce the predictions of QM, and likewise, KS theorem that no NCHVTs can reproduce the predictions of QM.
Bell's theorem can be proven either by the violation of a Bell inequality \cite{Bell64,CHSH} or by a logical contradiction \cite{GHZ89, TYO13, Hardy92, Hardy93, Cabello01}. KS theorem is usually proven by a logical contradiction \cite{KS67,Peres90,Peres91,Mermin90-KS,Mermin93-KS,Cabello96} or by a specific type of noncontextual inequalities~\cite{YO12,TY17}.
As for experimental respects, one would prefer to test Bell inequalities \cite{CHSH,Mermin-ineq}, noncontextual inequalities~\cite{Cabello08} and other alternative formulations even not directly involving the KS theorem \cite{PhysRevA.71.052108,tollaksen2007pre,PhysRevLett.113.200401,PhysRevLett.116.180401,PhysRevLett.120.040602,PRXQuantum.2.010331}.

Quantum entanglement and space-like separation are two crucial requirements for proof of Bell's theorem but are not necessary for KS theorem (the latter might be valid for any quantum systems with dimension $d\geq3$) \cite{PhysRevA.59.1070,PhysRevResearch.2.012068,PhysRevA.71.052108}.
We can still use these inequalities, which have the same mathematical form with Bell inequalities, to prove quantum contextuality~\cite{Karimi2010}. 
The inequalities are called Bell-like noncontextual inequalities.

Cabello proposed a 2-qubit state-dependent Bell-like noncontextual inequality {\cite{Cabello-FRH}} derived from the Peres-Mermin proof {\cite{Peres90,Peres91,Mermin90-KS,Mermin93-KS}} of KS theorem, which can be simplified by abandoning some items holding both for QM and NCHVTs.
Similar to the 3-party Bell inequality (Mermin inequality)~{\cite{Mermin-ineq}}, a 3-qubit state-dependent noncontextual inequality can also be worked out from the Mermin's proof of KS theorem~{\cite{Mermin90-KS,Mermin93-KS}}.
It is referred to as the Mermin-like inequality.
Furthermore, by extending the assumption of ``elements of reality''~\cite{EPR} in NCHVTs on single observables to composite observables, one can establish several other Peres-Mermin-type proofs of KS theorem and the corresponding state-dependent noncontextual inequalities based on Greenberger-Horne-Zeilinger (GHZ) states in 3-qubit case~\cite{GHZ89,Cramer2019}.
Interestingly, the mathematical forms of these state-dependent noncontextual inequalities are same as the Mermin-like one up to some cancellable items holding for both QM and NCHVTs, which implies a special kind of the equivalent relations.

Like nonlocality correlation of QM, quantum contextuality can be used as a resource \cite{2017ContextualityResource} for many quantum tasks such as simulation of quantum information processes, quantum error correction~\cite{1997code}, random access codes~\cite{galvao2002foundations}, one-location quantum games~\cite{2008Aharon}, and universal quantum computing \cite{PhysRevLett.102.010401,howard2014contextuality,hoban2011non}.
In addition, contextuality has been proved as a remarkable equivalent to the possibility of universal quantum computation via ``magic state'' distillation~\cite{2005Kitaev,howard2014contextuality}.
In this paper, we study several typical equivalent noncontextual inequalities from different noncontextual models of the proofs for Kochen-Specker theorem.
A single-photon GHZ entangled state with spin angular momentum (SAM), orbital angular momentum (OAM) and path degrees of freedom (DoFs) is prepared in experiment.
We test all these state-dependent noncontextual models by checking an extreme simplest Mermin-like inequality.
The experimental results show that the Mermin-like inequality can be obviously violated and quantum mechanics is contextual.

\section{Theory}\label{sectheory}
Some proofs of KS theorems in 3-qubit case exhibit an interesting kind of equivalent relations, namely, they can lead to the same noncontextual inequality up to the cancellable items.
Here, we consider several equivalent state-dependent proofs of KS theorem based on a special kind of multi-DoF (Path, SAM and OAM) GHZ states of a single photon.
The first GHZ-type state-dependent proof of KS theorem comes from Mermin {\cite{Mermin90-KS, Mermin93-KS}}. The main idea is as follow.

Begin with marking quantum states of a single photon with different DoFs.
For path DoF, eigenstates of measurement $Z^{\mathrm{p}}$ have eigenvalues $+1$ and $-1$ for $\left|{0}\right>^{\mathrm{p}}$ path mode and $\left|{1}\right>^{\mathrm{p}}$ path mode, respectively.
Similar to path DoF, $\left|{0}\right>^{\mathrm{s}}$ and $\left|{1}\right>^{\mathrm{s}}$ are horizontal and vertical polarization states of SAM, and $\left|{0}\right>^{\mathrm{o}}$ and $\left|{1}\right>^{\mathrm{o}}$ are OAM states with different OAM values $\ell=+1$ and $\ell=-1$, respectively.
Besides, $\left|{\pm}\right>=(\left|{0}\right>\pm\left|{1}\right>)/\sqrt{2}$ and $\left|{R(L)}\right>=(\left|{0}\right>\pm \mathrm{i}\left|{1}\right>)/\sqrt{2}$.
Therefore, the multi-DoF GHZ state of a single photon can be written as
\begin{equation}\label{eqGHZ}
\left|{G}\right>=(\left|{0}\right>^{\mathrm{o}}\left|{0}\right>^{\mathrm{s}}\left|{0}\right>^{\mathrm{p}}+\left|{1}\right>^{\mathrm{o}}\left|{1}\right>^{\mathrm{s}}\left|{1}\right>^{\mathrm{p}})/\sqrt{2}.
\end{equation}

Note that measurements carried on different DoFs of the multi-DoF GHZ state can no longer be considered as space-like separated events.
Therefore, a conventional GHZ test to demonstrate nonlocal quantum correlations using Bell inequalities cannot be directly adopted here.
Alternatively, we can modify the experimental protocol to a weaker version, i.e., a GHZ-like test due to Mermin to reveal the conflict between NCHVTs and QM.

\begin{figure}[thbp]
  \centering
  \includegraphics[scale=0.8]{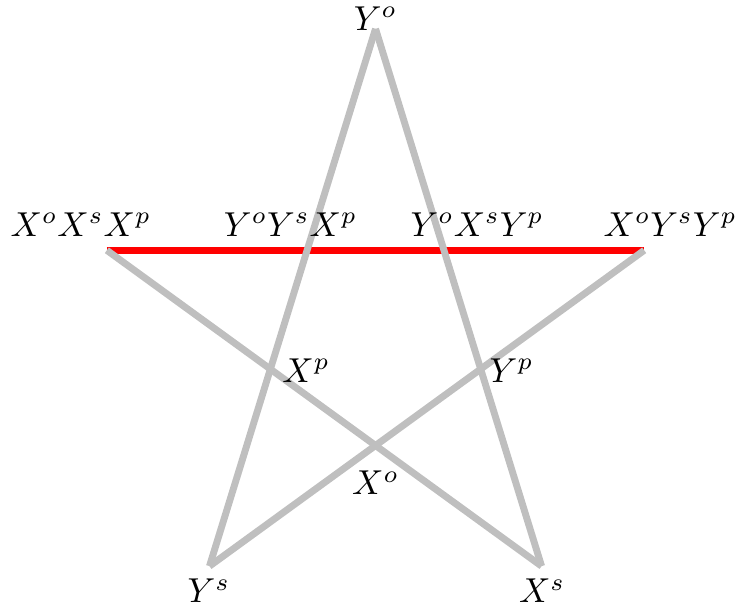}
  \caption{Mermin's pentagram proof for KS theorem. The observables along each line are in the same context (mutually commuting). The product of the four observables on every gray line of the star is $I$. In contrast, the product of the observables on red line is $-I$. Therefore, any noncontextual value assignment to each observable will lead to a contradiction.}\label{Merminpentagram}
\end{figure}

Consider the GHZ state $\left|{G}\right>$.
The GHZ-type proof of KS theorem can be derived using the four contexts $\{X^{\mathrm{o}}, Y^{\mathrm{s}}, Y^{\mathrm{p}}\}$, $\{Y^{\mathrm{o}}, X^{\mathrm{s}}, Y^{\mathrm{p}}\}$, $\{Y^{\mathrm{o}}, Y^{\mathrm{s}}, X^{\mathrm{p}}\}$  and $\{X^{\mathrm{o}}, X^{\mathrm{s}}, X^{\mathrm{p}}\}$ (displayed in Fig.~\ref{Merminpentagram}).
The following four QM predictions for the contexts in the GHZ state $\left|{G}\right>$ are
\begin{subequations}\label{QM}
\begin{align}
X^{\mathrm{o}}\cdot Y^{\mathrm{s}}\cdot Y^{\mathrm{p}}\left|{G}\right>&=-\left|{G}\right>,\label{QM1}\\
Y^{\mathrm{o}}\cdot X^{\mathrm{s}}\cdot Y^{\mathrm{p}}\left|{G}\right>&=-\left|{G}\right>,\label{QM2}\\
Y^{\mathrm{o}}\cdot Y^{\mathrm{s}}\cdot X^{\mathrm{p}}\left|{G}\right>&=-\left|{G}\right>,\label{QM3}\\
X^{\mathrm{o}}\cdot X^{\mathrm{s}}\cdot X^{\mathrm{p}}\left|{G}\right>&=+\left|{G}\right>.\label{QM4}
\end{align}
\end{subequations}
Since the measured results of $\{X^{\mathrm{\alpha}},Y^{\mathrm{\alpha}},\alpha\in\{o,s,p\}\}$ must be the eigenvalues $\pm1$,
according to the noncontextual assignment assumption of NCHVTs,  the involved six observables $X^{\mathrm{o}}$, $X^{\mathrm{s}}$, $X^{\mathrm{p}}$, $Y^{\mathrm{o}}$, $Y^{\mathrm{s}}$ and $Y^{\mathrm{p}}$ should be assigned to predefined values $v(X^{\mathrm{o}})$, $v(X^{\mathrm{s}})$, $v(X^{\mathrm{p}})$, $v(Y^{\mathrm{o}})$, $v(Y^{\mathrm{s}})$, and $v(Y^{\mathrm{p}})$, respectively, where $v(A^{\alpha})=\pm1$, and $A\in\{X,Y\}$.
It follows that if the observables from one context, namely, mutually commuting, satisfying a certain algebraic relation, then the values assigned to them in an individual system must obey the algebraic constraint of the same structure.
For these predefined values, one has the following constraints from Eqs. (\ref{QM1})-(\ref{QM4}),
\begin{subequations}\label{KMGHZ}
\begin{align}
v(X^{\mathrm{o}})\cdot v(Y^{\mathrm{s}})\cdot v(Y^{\mathrm{p}})&=-1,\label{KMGHZ1}\\
v(Y^{\mathrm{o}})\cdot v(X^{\mathrm{s}})\cdot v(Y^{\mathrm{p}})&=-1,\label{KMGHZ2}\\
v(Y^{\mathrm{o}})\cdot v(Y^{\mathrm{s}})\cdot v(X^{\mathrm{p}})&=-1,\label{KMGHZ3}\\
v(X^{\mathrm{o}})\cdot v(X^{\mathrm{s}})\cdot v(X^{\mathrm{p}})&=+1.\label{KMGHZ4}
\end{align}
\end{subequations}
According to the above discussion, $v^2(A^{\alpha})=1$. Then multiplying both sides of Eqs. (\ref{KMGHZ1})-(\ref{KMGHZ4}), we find that the left-hand side is $v^2(X^{\mathrm{o}})v^2(Y^{\mathrm{o}})\cdot v^2(X^{\mathrm{s}})v^2(Y^{\mathrm{s}})\cdot v^2(X^{\mathrm{p}})v^2(Y^{\mathrm{p}})=+1$, which contradicts the product of the four right-hand side numbers $(-1)^3\cdot(+1)=-1$. Therefore, it is impossible to ascribe predefined values $-1$ or $+1$ to each of the six observables.

In fact, the algebraic structure used in the above argument is the same as that in the 3-qubit GHZ paradox except for their physical explanations.
Any proof of KS theorem can be converted to an experimentally testable noncontextual inequality~\cite{Cabello08,HuXiaoMin2016}. Here the related noncontextual inequality can be derived from the linear combination of the expectation values of products of the variables in each of the contexts, with the respective quantum mechanical predictions as coefficients.
If each of the six variables can be assigned to predefined values $-1$ or $+1$, then we can get the following Mermin-like inequality \cite{PhysRevLett.65.1838,PhysRevA.65.012107}
\begin{align}\label{equality1}
&-\langle X^{\mathrm{o}}\cdot Y^{\mathrm{s}}\cdot Y^{\mathrm{p}}\rangle-\langle Y^{\mathrm{o}}\cdot X^{\mathrm{s}}\cdot  Y^{\mathrm{p}}\rangle\notag\\
&-\langle Y^{\mathrm{o}}\cdot Y^{\mathrm{s}} \cdot X^{\mathrm{p}}\rangle+\langle X^{\mathrm{o}}{\cdot}X^{\mathrm{s}}{\cdot}X^{\mathrm{p}}\rangle\leq2
\end{align}
and we have a brief proof for the classical bound in Appendix \ref{bound}.
As shown by Eqs. (\ref{QM1})-(\ref{QM4}), however, quantum mechanics predicts the maximal value of the combination to be 4 (with ideal equipment). In fact, the actual quantum violation measured in experiment is always less than 4 since the noise of environment and imperfection of experimental equipment.

The inequality (\ref{equality1}) is state-independent. This can be proved by checking a deep connection between the above GHZ-type proof of KS theorem and the Mermin's pentagram proof \cite{Mermin93-KS} shown in Fig. \ref{Merminpentagram}.
Note that the involved items in the GHZ-type proof, say $X^{\mathrm{o}}Y^{\mathrm{s}}Y^{\mathrm{p}}$, $Y^{\mathrm{o}}X^{\mathrm{s}}Y^{\mathrm{p}}$, $Y^{\mathrm{o}}Y^{\mathrm{s}}X^{\mathrm{p}}$ and $X^{\mathrm{o}}X^{\mathrm{s}}X^{\mathrm{p}}$, are exactly the products of the single qubit observables in the remained four lines.
One can infer (although not very strictly) that the GHZ-type proof is state-dependent from the fact that the six single qubit observables generate a subset of the Mermin's pentagram set (made up of ten observables in Fig. \ref{Merminpentagram}) while the whole pentagram set can provide a state-independent proof for KS theorem.

In above, we only consider the form of independent measurements with DoFs.
As for other forms, we discuss several typical proofs of KS theorem and the corresponding noncontextual inequalities in Appendix \ref{app1}. 
Following the similar arguments to Ref. {\cite{Cabello-FRH}}, since the cancellable items hold both for QM and NCHVTs and we use a multi-stage (cascade) measurement method (for example, the experiment result of $\left<X^\mathrm{s}\cdot Y^{\mathrm{o}}Y^{\mathrm{p}}\right>$ is the same as $\left<X^\mathrm{s}\cdot Y^{\mathrm{o}}\cdot Y^{\mathrm{p}}\right>$),
we can use the same experimental setup to test the same simplified inequality in the form of Eq. (\ref{equality1}) in principle.
Furthermore, we discuss a framework to study the commonness of the GHZ-Peres-Mermin type proofs of KS theorem for $n$ qubits in Appendix \ref{app-g}.

\section{Experimental setup and results}
\begin{figure}[t]
  \centering
  \includegraphics[width=0.95\linewidth]{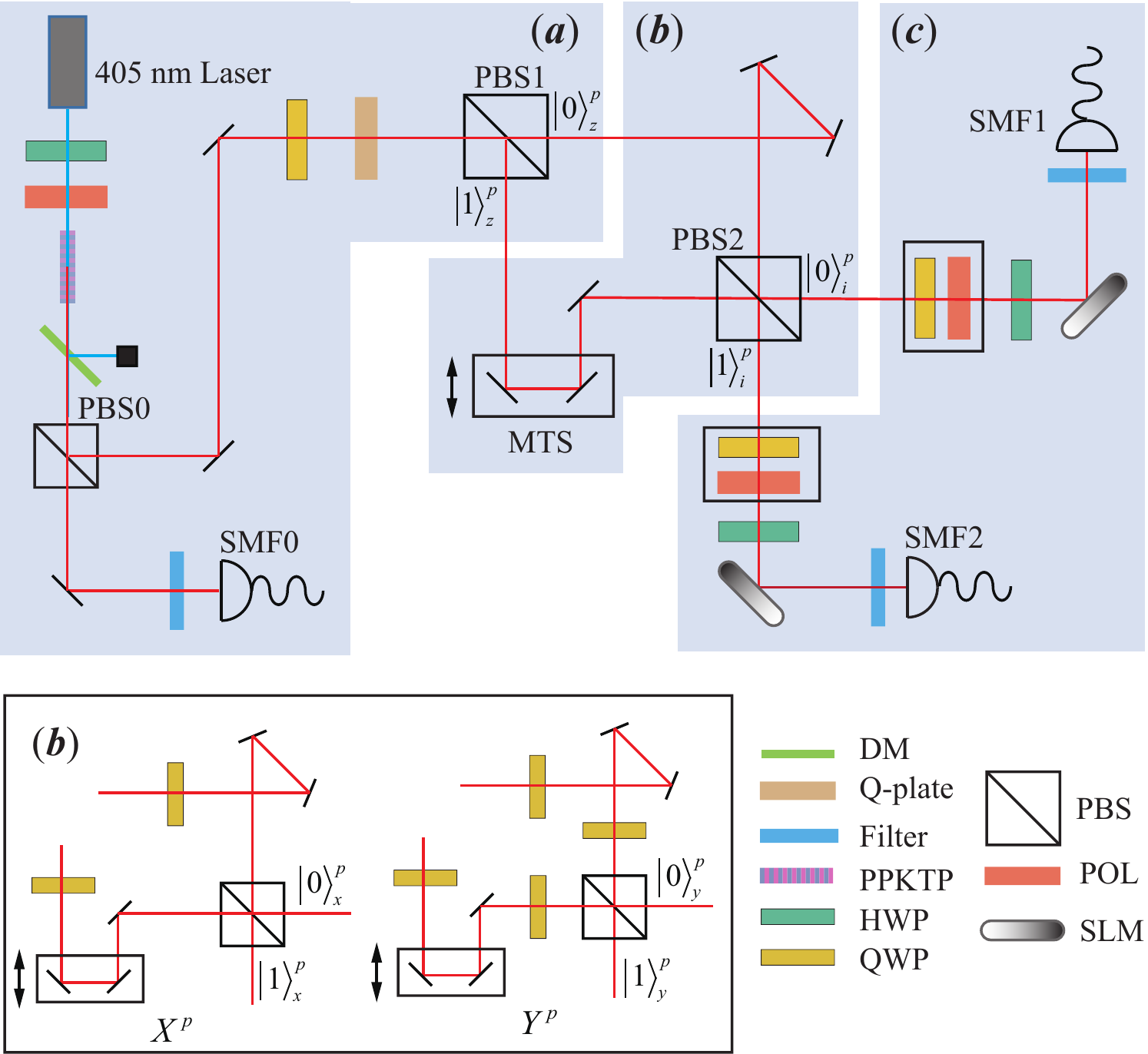}
  \caption{Experimental setup for GHZ tests with a single photon. (a) Initial state preparation. A single mode beam with a central wavelength of $405$ nm is focus on a $5$ mm ppKTP nonlinear crystal to create correlated photon pairs at $810$ nm. QWP, Q-plate, and PBS are used to generated the GHZ state between SAM, OAM, and path DoFs. (b)-(c) Realization of the projection measurements of path, SAM and radial OAM, correspond to a Mach-Zehnder interferometer, a combination of QWP and POL, and SLM together with a SMF, respectively. ppKTP: periodically poled potassium titanyl phosphate, PBS: polarizing beam splitter, HWP: half-wave plate, QWP: quarter-wave plate, DM: Dichroic mirror, MTS: motorized translation stage, POL: polarizer, SLM: spatial light modulator, SMF: single mode fiber.}\label{fig.exp}
\end{figure}

Our experimental setup is shown in Fig.~\ref{fig.exp}.
The generation of single-photon GHZ states is illustrated in Fig.~\ref{fig.exp}(a).
Correlated photon pairs in the SAM basis $\{\left|{0}\right>^{\mathrm{s}}, \left|{1}\right>^{\mathrm{s}}\}$ are created through a type-\Rmnum{2} spontaneous parametric down-conversion (SPDC) in a $5$-mm-long periodically poled potassium titanyl phosphate (ppKTP) crystal pumped by a $405$ nm continuous-wave diode laser.
The dichroic mirror is to make the $810$ nm correlated photons transmit, and the $405$ nm pump light is reflected into the block box.
The correlated photon pairs are separated through a polarizing beam splitter (PBS), and the horizontally polarized photons are coupled into the single mode fiber (SMF0) as a trigger signal.

Then only need to consider the vertical polarizing photon, the initial state is expressed as $\left|{\psi}\right>=\left|{\ell=0}\right>\left|{1}\right>^{\mathrm{s}}\left|{0}\right>^{\mathrm{p}}$.
The quarter wave plate (QWP) after PBS0 turns the vertical polarization to left-handed polarization, $\left|{\psi}\right>=\left|{\ell=0}\right>(\left|{0}\right>^{\mathrm{s}}+\mathrm{i}\left|{1}\right>^{\mathrm{s}})\left|{0}\right>^{\mathrm{p}}/\sqrt{2}$.
The quantum states of SAM and OAM coupling can be generated by $q$-plate, which is a kind of phase plate with locally varying birefringence that give rise to such coupling through the Pancharatnam-Berry geometric phase \cite{PhysRevLett.96.163905}.
The unitary operation of $q$-plate ${U}_{q}$ can be defined as
${U}_{q}\left|{L,\ell}\right>=\mathrm{i}\left|{R,\ell+2q}\right>$ and ${U}_{q}\left|{R,\ell}\right>=\mathrm{i}\left|{L,\ell-2q}\right>$,
in which $q$ is the topological charge of the $q$-plate, here $q=1/2$ in our experiment and L (R) represents left (right)-handed polarization.
Therefore, it changes to the state $\left|{\psi}\right>=\left|{0}\right>^{\mathrm{o}}(\left|{0}\right>^{\mathrm{s}}-\mathrm{i}\left|{1}\right>^{\mathrm{s}})\left|{0}\right>^{\mathrm{p}}/\sqrt{2}$ after the $q$-plate.
The incident photons pass through PBS that separates the photons into the $\left|{0}\right>^{\mathrm{p}}$ path mode and the $\left|{1}\right>^{\mathrm{p}}$ path mode, meanwhile, the OAM value of down path flips from $+\hbar$ to $-\hbar$ after the reflection of PBS. The multi-DoF GHZ state $\left|{G}\right>$ with single photons has been generated successfully after passing through PBS.

The projective measurements of path DoF $X^{\mathrm{p}}$ and $Y^{\mathrm{p}}$ are illustrated in Fig.~\ref{fig.exp}(b).
The measurement $X^{\mathrm{p}}$, inserting the QWPs (all setting in $45^{\circ}$) into the two optical paths in the Mach-Zehnder interferometer respectively, is $+1(-1)$ for the $\left|{+}\right>^{\mathrm{p}}$($\left|{-}\right>^{\mathrm{p}}$) path mode through the PBS.
Considering the inversion of the OAM modes caused by reflection, we revise the configuration of interferometer to ensure the OAM modes of the two arms matching with projection measurements.
The difference between measurement $Y^{\mathrm{p}}$ and $X^{\mathrm{p}}$ is that another two QWPs (all setting in $90^{\circ}$) are added into the interferometer shown in the Fig.~\ref{fig.exp}(b), respectively.
The two PBSs here are equivalent to achieving global operations in these three DoFs.
The polarizers oriented at $45^{\circ}$ and QWPs depicted in Fig.~\ref{fig.exp}(c) allow measurements of linear polarization $\left|{+}\right>^{\mathrm{s}}, \left|{-}\right>^{\mathrm{s}}$ (circular polarization $\left|{R}\right>^{\mathrm{s}}, \left|{L}\right>^{\mathrm{s}}$).
The following HWPs after polarizers convert linear polarized photons to horizontal, which mode can be modulated by a spatial light modulator (SLM).
Finally a SLM and a SMF are used to perform any directional projection measurements of OAM modes.
An incoming photon is flattened its phase by SLM, and transformed into a Gaussian mode that can be efficiently coupled into the SMF.
Three single-photon avalanche detectors recorded photon counting rate after the multi-DoF Pauli measurements, and their coincidence counts are proportional to the detecting correlated photon pairs with a certain setting.

In addition, observables such like $Y^{\mathrm{o}}\cdot Y^{\mathrm{p}}$ and $Y^{\mathrm{o}}Y^{\mathrm{p}}$, scilicet the assigned predefined values $v(Y^{\mathrm{o}})\cdot v(Y^{\mathrm{p}})$ and $v(Y^{\mathrm{o}}Y^{\mathrm{p}})$, can be measured with a same experimental setup.
Furthermore, as described in Ref.~\cite{Cabello-FRH}, in the experiment we designed, these observables whose correspond predictions hold in any NCHVT can be measured by cascading the above-mentioned measurement methods.

To evaluate the performance of the GHZ state generated in experiment, the fidelity between the theoretical GHZ state $\left|{G}\right>$ and the prepared state $\rho_{\mathrm{exp}}$ is $\bar{F}_{\mathrm{exp}}=\left<G\right|\rho_{\mathrm{exp}}\left|{G}\right>=(90.0\pm3.0)\%$.
From this value, the entanglement witness of the GHZ state, defined by $\mathcal{W}=\mathcal{I}/2-\left|{G}\right>\left<{G}\right|$ \cite{Witness2004}, can be directly determined as $F_{\mathrm{wit}}=-0.400\pm0.030$. 
It is negative and thus proves the presence of genuine tripartite entanglement in our experiment, which means that this three-qubit state cannot be divided into any separated parts, and is indeed genuinely tripartite GHZ entangled.
All the experimental data are listed in Appendix \ref{app2}.

As explained above, demonstrations of the contradiction between NCHVTs and QM are exactly the four experiments ($xyy$, $yxy$, $yyx$ and $xxx$) with GHZ argument.
For each DoF, recall these elements of reality $X^i$ with values $+1(-1)$ for corresponding eigenstates $\left|{+}\right>^{\mathrm{i}}(\left|{-}\right>^{\mathrm{i}})$ in NCHVTs, similarly to $Y^i$.
Firstly we perform three experiments ($xyy$, $yxy$, and $yyx$), and the expectation values of three experiments measured are listed in Appendix \ref{app2}, $\left<X^{\mathrm{o}}Y^{\mathrm{s}}Y^{\mathrm{p}}\right>=-0.863\pm0.028$, $\left<Y^{\mathrm{o}}X^{\mathrm{s}}Y^{\mathrm{p}}\right>=-0.869\pm0.034$, and $\left<Y^{\mathrm{o}}Y^{\mathrm{s}}X^{\mathrm{p}}\right>=-0.897\pm0.032$.
\begin{figure}[t]
  \centering
  \includegraphics[width=\linewidth]{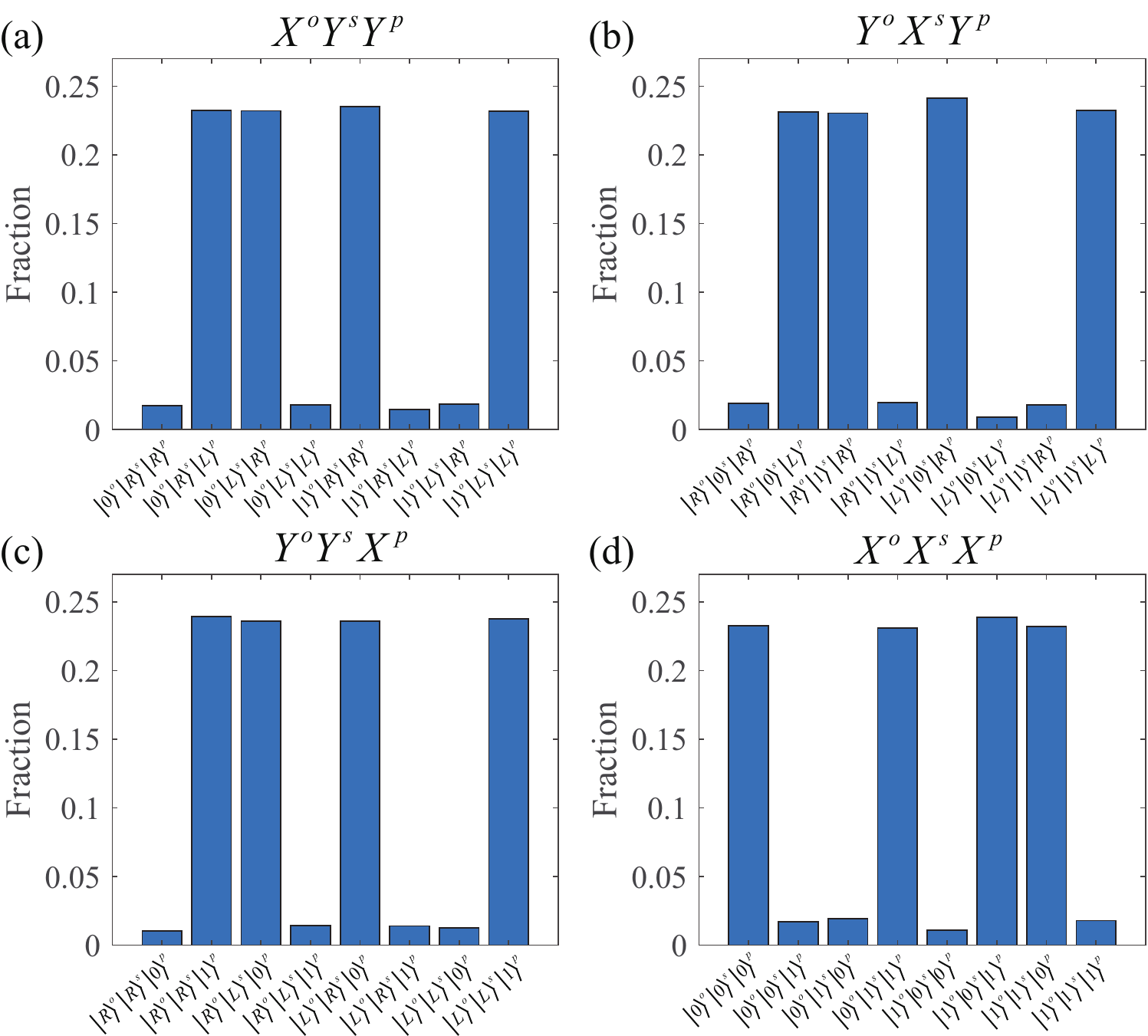}
  \caption{Data showing the fraction of measurements (a) $X^{\mathrm{o}}Y^{\mathrm{s}}Y^{\mathrm{p}}$, (b) $Y^{\mathrm{o}}X^{\mathrm{s}}Y^{\mathrm{p}}$, (c) $Y^{\mathrm{o}}Y^{\mathrm{s}}X^{\mathrm{p}}$ and (d) $X^{\mathrm{o}}X^{\mathrm{s}}X^{\mathrm{p}}$ on the corresponding eigenstates base, respectively.}\label{data1}
\end{figure}
It supports that the perfect correlations, i.e. Eqs. (\ref{KMGHZ1})–(\ref{KMGHZ3}), cannot be easily obtained in real experiments.
The specific measurement results, illustrated in the Fig. \ref{data1}(a)-\ref{data1}(c), show the probabilities of the three experiments roughly analogous in respective bases, which are in line with theory.

Then consider the fourth experiment $xxx$.
On the one hand, we assume that multiplication rules are valid for NCHVTs, and any measurements between different DoFs must be independent in NCHVTs.
One can obtain $v(X^{\mathrm{o}}){\cdot}v(X^{\mathrm{s}}){\cdot}v(X^{\mathrm{p}})=[v(X^{\mathrm{o}})\cdot v(Y^{\mathrm{s}})\cdot v(Y^{\mathrm{p}})]$$\cdot$$[v(Y^{\mathrm{o}})\cdot (X^{\mathrm{s}})\cdot v(Y^{\mathrm{p}})]\cdot[v(Y^{\mathrm{o}})\cdot v(Y^{\mathrm{s}})\cdot v(X^{\mathrm{p}})]=-1$.
It is easy to investigate the possible outcomes, that are predicted by noncontextual realism based on the elements of reality introduced to explain the above three $xyy$, $yxy$, and $yyx$ experiments.
Therefore the only possible outcomes of experiment $xxx$, still being similarly to the three experiments above, are $\left|{+}\right>^{\mathrm{o}}\left|{+}\right>^{\mathrm{s}}\left|{-}\right>^{\mathrm{p}}$, $\left|{+}\right>^{\mathrm{o}}\left|{-}\right>^{\mathrm{s}}\left|{+}\right>^{\mathrm{p}}$, $\left|{-}\right>^{\mathrm{o}}\left|{+}\right>^{\mathrm{s}}\left|{+}\right>^{\mathrm{p}}$ and $\left|{-}\right>^{\mathrm{o}}\left|{-}\right>^{\mathrm{s}}\left|{-}\right>^{\mathrm{p}}$ from NCHVTs.
The experimental results (illustrated in Fig. \ref{data1}(d)), however, show that the outcomes of the fourth are completely inconsistent with the prediction of NCHVTs.
The value of observable $X^{\mathrm{o}}X^{\mathrm{s}}X^{\mathrm{p}}$ in the forth experiment yields $0.869\pm0.034$, being approximately equal to $1$ from the prediction of QM (Eq.(\ref{KMGHZ4})).

On the other hand, without requiring the perfect correlations [Eqs.(\ref{KMGHZ1})-(\ref{KMGHZ4})] and assuming about multiplication rules, it is sufficient to have the inequality (\ref{equality1}) test, which means an experimentally testable state-independent quantum contextuality \cite{Cabello08}.
The expectation value of the Mermin operator in our experiments can be obtained as $\left<M\right>$$=\left<X^{\mathrm{o}}X^{\mathrm{s}}X^{\mathrm{p}}\right>-\left<X^{\mathrm{o}}Y^{\mathrm{s}}Y^{\mathrm{p}}\right>-\left<Y^{\mathrm{o}}X^{\mathrm{s}}Y^{\mathrm{p}}\right>-\left<Y^{\mathrm{o}}Y^{\mathrm{s}}X^{\mathrm{p}}\right>=3.498\pm0.130$.
This value is greater than the maximum predicted under the NCHVTs background.
Combined with our theoretical proof, this kind of test actually is an experimental demonstration of the GHZ paradox in a contextual manner.

\section{Conclusion}
In summary, we find an equivalent class of noncontextual models for the state-dependent proofs of KS theorem and the corresponding equivalent noncontextual inequalities, which can be converted into a simple Mermin-like inequality by canceling out the trivial items. Since the trivial items hold both for any NCHVTs and QM, we can simultaneously check all these noncontextual models in the same experiment by examining the Mermin-like inequality.
In our experiment, SAM, OAM and path of a single photon are prepared in the GHZ state and the violation of the Mermin-like inequality exclusively discards any NCHVTs as a possible extension to QM.

A number of open questions can be raised in further research.
One of the interesting problems is to classify different noncontextual proofs of KS theorem for a larger number of qubits. Besides, the intrinsic causes of this classification need further study.
We conjecture that one of the causes may depend on whether the whole  information (predefined values of the respective observables) for the composite subsystems in each noncontextual models can be divided into smaller units of determined information, e.g., the information of the composite spin and OAM qubits in the models of Fig. \ref{PM-gauge4}(d) in Appendix \ref{app1} can only be considered as a whole.
Furthermore, the contextual advantage can be used for state-dependent cloning \cite{Lostaglio2020contextualadvantage,Sainz2020whatisnonclassical}, state discrimination \cite{PhysRevX.8.011015} and other quantum information tasks \cite{PhysRevResearch.2.023029}.
Since the universality for quantum computation schemes by state injection in Ref. \cite{2017ContextualityResource} tells us that any quantum unitary operations can be linked to some state-dependent proofs for quantum contextuality from a quantum resource prospective, then conversely, can different equivalent noncontextual proofs (not limited to single-qubit operation) be used to realize an universal  quantum gate?
Other questions such as how to construct a equivalent noncontextual models and inequalities from a rank-one projector set may also be interesting tasks.

\begin{acknowledgments}
S. Ru and W. Tang contributed equally to this work.
This work was supported by the National Nature Science Foundation of China (Grant Nos. 12074307, 11804271 and 91736104) and China Postdoctoral Science Foundation via Project No. 2020M673366.
\end{acknowledgments}

\appendix

\section{Classical bound for Mermin inequality}\label{bound}
This section is about a brief proof of classical bound for Mermin inequality. Let $a=v(X^{\mathrm{o}})\cdot v(Y^{\mathrm{s}})\cdot v(Y^{\mathrm{p}})$,
$b=v(Y^{\mathrm{o}})\cdot v(X^{\mathrm{s}})\cdot v(Y^{\mathrm{p}})$,
$c=v(Y^{\mathrm{o}})\cdot v(Y^{\mathrm{s}})\cdot v(X^{\mathrm{p}})$, and $abc=v(X^{\mathrm{o}})\cdot v(X^{\mathrm{s}})\cdot v(X^{\mathrm{p}})$. Here $\left|a\right|=1$, $\left|b\right|=1$, $\left|c\right|=1$, and we can have
\begin{align}
&(a+b+c-abc)^2=a^2+b^2+c^2+(abc)^2\notag\\
&\quad\quad+2ab(1-c^2)+2bc(1-a^2)+2ac(1-b^2)\notag\\
&\quad\quad= a^2+b^2+c^2+(abc)^2= 4.
\end{align}
Therefore, $a+b+c-abc=\pm2$, which means that the classical up bound of Mermin inequality is $+2$.

\section{Peres-Mermin square proofs}\label{app1}

\begin{figure}[t]
  \centering
  \includegraphics[width=\linewidth]{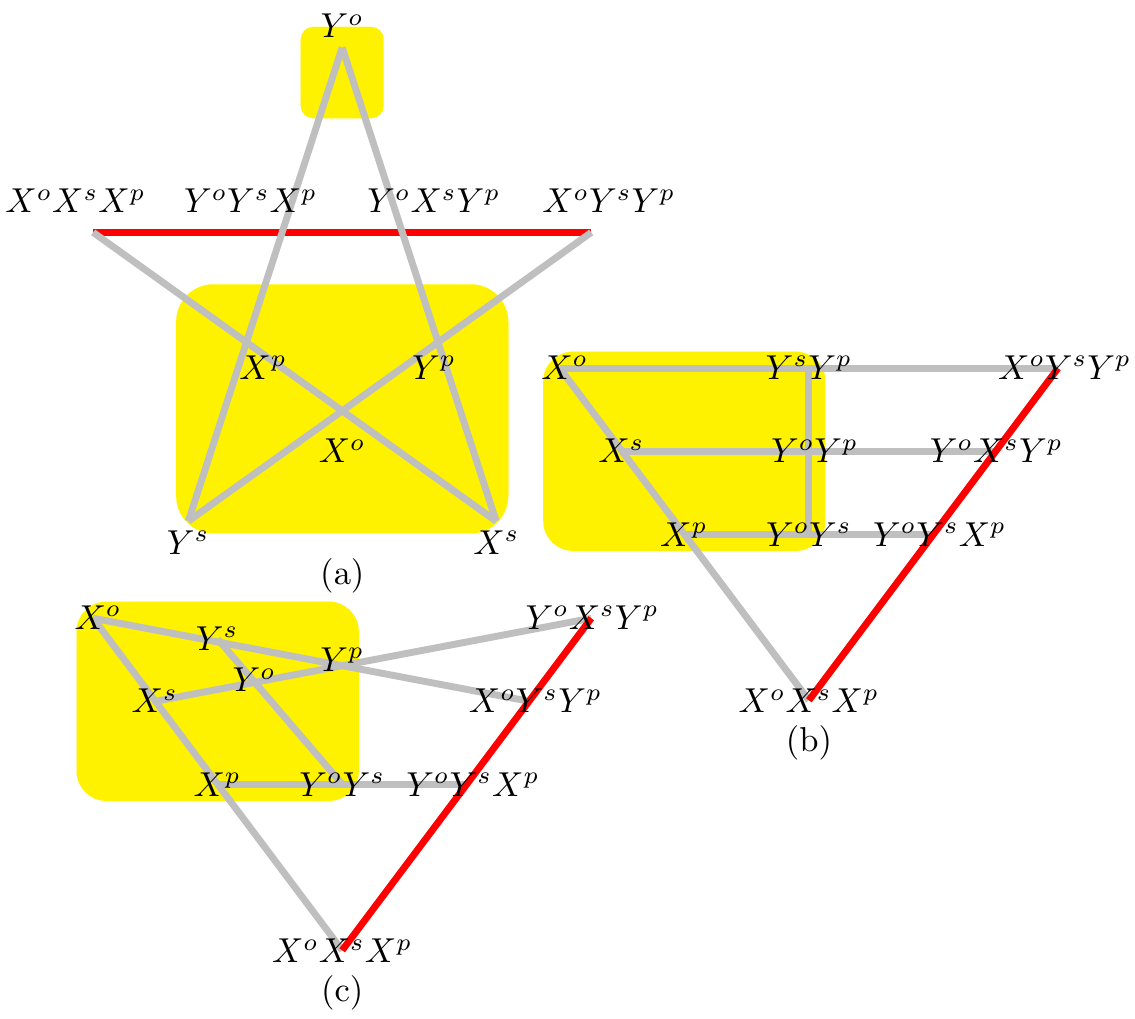}
  \caption{Three 3-qubit noncontextual models provide some representative  GHZ-Peres-Mermin type proofs  for KS theorem. Observables along each line are in the same context. The product of the observables on every gray line is $I$, but the red line is $-I$. Therefore, any noncontextual value assignment ($+1$ or $-1$) to each observable will lead to a contradiction. Besides, in each model, observables in the yellow zone (a subset) can provide a state-dependent proofs for KS theorem (by choosing the GHZ state $\left|{G}\right>$).}\label{PM-gauge4}
\end{figure}

Figure~\ref{PM-gauge4} shows three representative Peres-Mermin square proofs of KS theorem in which, Fig.~\ref{PM-gauge4}(a) are discussed in Sec. \ref{sectheory}.
\begin{table*}[!htbp]
\newcommand{\tabincell}[2]{\begin{tabular}{@{}#1@{}}#2\end{tabular}}
\begin{tabular}{ccc}
  \hline
  \hline
  Figure & value assignment constraints &  noncontextual inequalities \\
    \hline
  (a) & \tabincell{c}{$-v(X^{\mathrm{o}})\cdot v(Y^{\mathrm{s}})\cdot v(Y^{\mathrm{p}})=+1,$\\
$-v(Y^{\mathrm{o}})\cdot v(X^{\mathrm{s}})\cdot v(Y^{\mathrm{p}})=+1,$\\
$-v(Y^{\mathrm{o}})\cdot v(Y^{\mathrm{s}})\cdot v(X^{\mathrm{p}})=+1,$\\
$v(X^{\mathrm{o}}){\cdot}v(X^{\mathrm{s}}){\cdot}v(X^{\mathrm{p}})=+1.$}& \tabincell{c}{$-\langle X^{\mathrm{o}}\cdot Y^{\mathrm{s}}\cdot Y^{\mathrm{p}}\rangle-\langle Y^{\mathrm{o}}\cdot X^{\mathrm{s}}\cdot Y^{\mathrm{p}}\rangle-\langle Y^{\mathrm{o}}\cdot Y^{\mathrm{s}}\cdot X^{\mathrm{p}}\rangle$\\
$+\langle X^{\mathrm{o}}{\cdot}X^{\mathrm{s}}{\cdot}X^{\mathrm{p}}\rangle\leq2$}\\
    \hline
(b)& \tabincell{c}{$-v(X^{\mathrm{o}})\cdot v(Y^{\mathrm{s}}Y^{\mathrm{p}})=+1,$\\
$-v(X^{\mathrm{s}})\cdot v(Y^{\mathrm{o}}Y^{\mathrm{p}})=+1,$\\
$-v(X^{\mathrm{p}})\cdot v(Y^{\mathrm{o}}Y^{\mathrm{s}})=+1,$\\
$v(X^{\mathrm{o}}){\cdot}v(X^{\mathrm{s}}){\cdot}v(X^{\mathrm{p}})=+1,$\\
$v(Y^{\mathrm{s}}Y^{\mathrm{p}}){\cdot}v(Y^{\mathrm{o}}Y^{\mathrm{p}}){\cdot}v(Y^{\mathrm{o}}Y^{\mathrm{s}})=+1. $}& \tabincell{c}{$-\langle X^{\mathrm{o}}\cdot Y^{\mathrm{s}}Y^{\mathrm{p}}\rangle-\langle X^{\mathrm{s}}\cdot Y^{\mathrm{o}}Y^{\mathrm{p}}\rangle-\langle X^{\mathrm{p}}\cdot Y^{\mathrm{o}}Y^{\mathrm{s}}\rangle$\\
$+\langle X^{\mathrm{o}}{\cdot}X^{\mathrm{s}}{\cdot}X^{\mathrm{p}}\rangle+\langle Y^{\mathrm{s}}Y^{\mathrm{p}}{\cdot}Y^{\mathrm{o}}Y^{\mathrm{p}}{\cdot}Y^{\mathrm{o}}Y^{\mathrm{s}}\rangle\leq3$}\\
    \hline
(c) & \tabincell{c}{
$-v(X^{\mathrm{o}})\cdot v(Y^{\mathrm{s}})\cdot v(Y^{\mathrm{p}})=+1,$\\
$-v(Y^{\mathrm{o}})\cdot v(X^{\mathrm{s}})\cdot v(Y^{\mathrm{p}})=+1,$\\
$-v(Y^{\mathrm{o}}Y^{\mathrm{s}})\cdot v(X^{\mathrm{p}})=+1,$\\
$v(X^{\mathrm{o}}){\cdot}v(X^{\mathrm{s}})\cdot v(X^{\mathrm{p}})=+1,$\\
$v(Y^{\mathrm{s}}){\cdot}v(Y^{\mathrm{o}})\cdot v(Y^{\mathrm{s}}Y^{\mathrm{o}})=+1.$} & \tabincell{c}{$-\langle X^{\mathrm{o}}\cdot Y^{\mathrm{s}}\cdot Y^{\mathrm{p}}\rangle-\langle Y^{\mathrm{o}}\cdot X^{\mathrm{s}}\cdot Y^{\mathrm{p}}\rangle-\langle Y^{\mathrm{o}} Y^{\mathrm{s}}\cdot X^{\mathrm{p}}\rangle$\\
$+\langle X^{\mathrm{o}}{\cdot}X^{\mathrm{s}}{\cdot}X^{\mathrm{p}}\rangle+\langle Y^{\mathrm{s}}{\cdot}Y^{\mathrm{o}}{\cdot}Y^{\mathrm{s}}Y^{\mathrm{o}}\rangle\leq3$} \\
    \hline
  \hline
\end{tabular}\caption{Constraints for value assignments and the respective noncontextual  inequalities for GHZ-Peres-Mermin type proofs for KS theorem in Fig.~\ref{PM-gauge4}.}\label{PM-gauge-table}
\end{table*}
As for Fig.~\ref{PM-gauge4}(b),
by applying the six observables $X^{\mathrm{o}}$, $X^{\mathrm{s}}$, $X^{\mathrm{p}}$, $Y^{\mathrm{s}}Y^{\mathrm{p}}$, $Y^{\mathrm{o}}Y^{\mathrm{p}}$ and $Y^{\mathrm{o}}Y^{\mathrm{s}}$ in the yellow box on the multi-DoF GHZ state $\left|{G}\right>$, we can infer that the predefined values $v(X^{\mathrm{o}})$, $v(X^{\mathrm{s}})$, $v(X^{\mathrm{p}})$, $v(Y^{\mathrm{s}}Y^{\mathrm{p}})$, $v(Y^{\mathrm{o}}Y^{\mathrm{p}})$, and $v(Y^{\mathrm{o}}Y^{\mathrm{s}})$ satisfy the following equations according to the assumptions of NCHVTs,
\begin{subequations}\label{KMGHZ-a}
\begin{align}
-v(X^{\mathrm{o}})\cdot v(Y^{\mathrm{s}}Y^{\mathrm{p}})&=+1,\label{KMGHZ1-a}\\
-v(X^{\mathrm{s}})\cdot v(Y^{\mathrm{o}}Y^{\mathrm{p}})&=+1,\label{KMGHZ2-a}\\
-v(X^{\mathrm{p}})\cdot v(Y^{\mathrm{o}}Y^{\mathrm{s}})&=+1,\label{KMGHZ3-a}\\
v(X^{\mathrm{o}}){\cdot}v(X^{\mathrm{s}}){\cdot}v(X^{\mathrm{p}})&=+1,\label{KMGHZ4-a}\\
v(Y^{\mathrm{s}}Y^{\mathrm{p}}){\cdot}v(Y^{\mathrm{o}}Y^{\mathrm{p}}){\cdot}v(Y^{\mathrm{o}}Y^{\mathrm{s}})&=+1.\label{KMGHZ5-a}
\end{align}
\end{subequations}
Eqs.~(\ref{KMGHZ1-a}-\ref{KMGHZ4-a}) follow from the algebraic relations for the stabilizers of the state $\left|{G}\right>$, and Eq.~(\ref{KMGHZ5-a}) follows from the fact that the product of three mutually commuting operators is identity.
Note that each of the six observables appears twice in the left-hand sides of Eqs.~(\ref{KMGHZ1-a})-\ref{KMGHZ5-a}). Therefore, it is impossible to pre-assign the values ($-1$ or $+1$) to them.
The contradiction comes from the fact that the product of the left-hand sides of Eqs.~(\ref{KMGHZ1-a})-\ref{KMGHZ5-a}) is $-1$ while the right-hand side is $+1$.

An experimental testable inequality for the proof of KS theorem can be derived from the linear combination of the five expectation values
of the products of the three observables in Fig.~\ref{PM-gauge4}(b), with the respective quantum mechanical predictions as coefficients,
\begin{align}\label{equality0-a}
&-\left<X^{\mathrm{o}}\cdot Y^{\mathrm{s}}Y^{\mathrm{p}}\right>-\left<X^{\mathrm{s}}\cdot Y^{\mathrm{o}}Y^{\mathrm{p}}\right>-\left<X^{\mathrm{p}}\cdot Y^{\mathrm{o}}Y^{\mathrm{s}}\right>\notag\\
&+\left<X^{\mathrm{o}}{\cdot}X^{\mathrm{s}}{\cdot}X^{\mathrm{p}}\right>+\left<Y^{\mathrm{s}}Y^{\mathrm{p}}{\cdot}Y^{\mathrm{o}}Y^{\mathrm{p}}{\cdot}Y^{\mathrm{o}}Y^{\mathrm{s}}\right>\leq3.
\end{align}
Note that these items are exactly the products of observables in each row or each column of the yellow box.
For the GHZ state, the prediction of QM is 5.

The order of mutually commuting observables can be exchanged because observables in different DoFs are compatible, Eq. (\ref{KMGHZ5-a}) is state-independent and holds in any NCHVT.
In other words, in any NCHVT, those observables whose product is $\mathrm{I}$ or $\mathrm{-I}$ have an expectation of $1$ or $-1$.
Therefore, Eq. (\ref{equality0-a}) can be simplified to
\begin{align}\label{equality1-a}
  &\left<X^{\mathrm{o}}X^{\mathrm{s}}X^{\mathrm{p}}\right>-\left<X^{\mathrm{o}}Y^{\mathrm{s}}Y^{\mathrm{p}}\right>\notag\\
  &\qquad-\left<Y^{\mathrm{o}}X^{\mathrm{s}}Y^{\mathrm{p}}\right>-\left<Y^{\mathrm{o}}Y^{\mathrm{s}}X^{\mathrm{p}}\right>\leq2,
\end{align}
which is exactly the same form of KS inequality as that in the main body, and the prediction of QM is $4$.
From this point of view, one can say that they are indeed equivalent.

Similar arguments can be applied to the other model Fig.-\ref{PM-gauge4}(c).
The constraints of the value assignments and the  noncontextual  inequalities  are listed in Table \ref{PM-gauge-table}.
It follows that the noncontextual inequalities with respect to Fig.~\ref{PM-gauge4}(a)-(c) are equivalent noncontextual inequalities.
In fact, the reduced Bell operators of Fig.~\ref{PM-gauge4}(a)-(c) are
\begin{align*}
\mB_a^R=&-X^{\mathrm{o}}\cdot Y^{\mathrm{s}}\cdot Y^{\mathrm{p}}- X^{\mathrm{s}}\cdot Y^{\mathrm{o}}\cdot Y^{\mathrm{p}}\\
&- X^{\mathrm{p}}\cdot Y^{\mathrm{o}}\cdot Y^{\mathrm{s}}+ X^{\mathrm{o}}{\cdot}X^{\mathrm{s}}{\cdot}X^{\mathrm{p}},\\
\mB_b^R=&-X^{\mathrm{o}}\cdot Y^{\mathrm{s}} Y^{\mathrm{p}}- X^{\mathrm{s}}\cdot Y^{\mathrm{o}} Y^{\mathrm{p}}\\
&- X^{\mathrm{p}}\cdot Y^{\mathrm{o}} Y^{\mathrm{s}}+ X^{\mathrm{o}}{\cdot}X^{\mathrm{s}}{\cdot}X^{\mathrm{p}},\\
\mB_c^R=&-X^{\mathrm{o}}\cdot Y^{\mathrm{s}}\cdot Y^{\mathrm{p}}- X^{\mathrm{s}}\cdot Y^{\mathrm{o}}\cdot Y^{\mathrm{p}}\\
&- X^{\mathrm{p}}\cdot Y^{\mathrm{o}} Y^{\mathrm{s}}+ X^{\mathrm{o}}{\cdot}X^{\mathrm{s}}{\cdot}X^{\mathrm{p}},
\end{align*}
respectively.

\section{\textit{n}-qubit GHZ-Peres-Mermin type proofs of KS theorem}\label{app-g}
\begin{table*}[!hbtp]
\centering
\begin{tabular}{ccccccccc}
\hline\hline
State&$\left|{000}\right>$&$\left|{001}\right>$&$\left|{010}\right>$&$\left|{001}\right>$&$\left|{100}\right>$&$\left|{101}\right>$&$\left|{110}\right>$&$\left|{111}\right>$\\
\hline
Probability&0.485&0.037&0.003&0.002&0.003&0.003&0.027&0.440\\
Error&0.014&0.006&0.000&0.000&0.000&0.000&0.004&0.012\\
\hline
\hline
Observables & $X^{\mathrm{o}}Y^{\mathrm{s}}Y^{\mathrm{p}}$ & $Y^{\mathrm{o}}Y^{\mathrm{s}}X^{\mathrm{p}}$ & $Y^{\mathrm{o}}X^{\mathrm{s}}Y^{\mathrm{p}}$ & $X^{\mathrm{o}}X^{\mathrm{s}}X^{\mathrm{p}}$ & $Y^{\mathrm{o}}X^{\mathrm{s}}X^{\mathrm{p}}$ & $X^{\mathrm{o}}X^{\mathrm{s}}Y^{\mathrm{p}}$ & $X^{\mathrm{o}}Y^{\mathrm{s}}X^{\mathrm{p}}$ & $Y^{\mathrm{o}}Y^{\mathrm{s}}Y^{\mathrm{p}}$ \\
\hline
Expectation value&$-0.863$&$-0.897$&$-0.869$&$0.869$&$-0.014$&$-0.016$&$0.0059$&$-0.006$\\
Error&$0.028$&$0.032$&$0.034$&$0.036$&$0.027$&$0.028$&$0.030$&$0.032$\\
\hline\hline
\end{tabular}
\caption{Measurements of fidelity and witness of the multi-DoF GHZ state. First three lines, the results of Pauli measurement $Z^{\mathrm{o}}Z^{\mathrm{s}}Z^{\mathrm{p}}$ are list. From $000$ to $111$: $\left|{0}\right>^{\mathrm{o}}\left|{0}\right>^{\mathrm{s}}\left|{0}\right>^{\mathrm{p}}$, $\left|{0}\right>^{\mathrm{o}}\left|{0}\right>^{\mathrm{s}}\left|{1}\right>^{\mathrm{p}}$, $\cdots$, $\left|{1}\right>^{\mathrm{o}}\left|{1}\right>^{\mathrm{s}}\left|{1}\right>^{\mathrm{p}}$. The expectation values of other eight observables are list in last three lines.}\label{table1}
\end{table*}

Inspired by the properties of the sequential measurements \cite{Cabello-FRH}, we here propose a framework to study the commonness of the GHZ-Peres-Mermin type proofs of KS theorem for $n$ qubits. Based on this, as we will see below, one may test different noncontextual models by the same experiment.

Denote by $\{\mC, |\varphi\rangle; \mM(\mB)\}$  an $n$-qubit noncontextual model for GHZ-Peres-Mermin type proof of KS theorem, where $\mC=\{\mC_1,\mC_2,...,\mC_s\}$, each $\mC_i$ is a context (a set of mutually commuting POs), and $\mM(\mB)$ is the induced noncontextual inequality.
Let $\mC_i=\{A_i^1,A_i^2,...,A_i^{|\mC_i|}\}$, where $A_i^{j}$ ($j=1,2,...,|\mC_i|$) the $j$-th PO in the context, then $\{\mC, |\varphi\rangle; \mM(\mB)\}$ is {\it non-reducible} if it can not produce a new state-dependent noncontextual model by reducing any PO in all the involved contexts.
To give a strict framework, we assume that the state-dependent noncontextual models discussed in this paper are all non-reducible.
Then for the model $\{\mC, |\varphi\rangle; \mM(\mB)\}$, one can get the following $s$ equations:
\begin{align*}
    A_i^1\cdot A_i^2\cdot ...\cdot A_i^{|\mC_i|}|\varphi\rangle=\alpha_i|\varphi\rangle,~(i=1,2,...,s),
\end{align*}
where $\alpha_i\in\{1,-1\}$. According to the assumptions of NCHVTs, if the observables from one
context are constrained by a certain algebraic relation, then the values assigned to them must obey the same algebraic constraint. Based on this, the following constraints for these predefined values should be satisfied:
\begin{align*}
    v(A_i^1)\cdot v(A_i^2)\cdot ...\cdot v(A_i^{|\mC_i|})=\alpha_i,~(i=1,2,...,s).
\end{align*}
Moreover, the Bell operator can be defined as $\mB\equiv\sum_{i=1}^s\alpha_i A_i^1\cdot A_i^2\cdot ...\cdot A_i^{|\mC_i|}$, then the noncontextual inequality reads
\begin{align}\label{Noncontextual-ineq}
    \mM(\mB):~\langle \mB\rangle\leq \beta_c(\mC),
\end{align}
where $\langle\cdot\rangle$ represents the classical (noncontextual) expectation and $\beta_c(\mC)$ is a constant which is bounded by the noncontextual value assignments of the related observables. One can easily check that the quantum violation of $\langle \mB\rangle_{QM}$ is larger than $\beta_c(\mC)$ (its maximum is usually obtained in $|\varphi\rangle$).

In fact, for some GHZ-Peres-Mermin type proofs,  the noncontextual inequalities tested in the experiments are not the $\mM(\mB)$-like versions but rather simplified ones, e.g., a 2-qubit Bell-like inequality in Ref.~\cite{Cabello-FRH}. This is because for a given $\{\mC, |\varphi\rangle; \mM(\mB)\}$, if the product of the observables in  some contexts $\mC_{l_1},...,\mC_{l_k}\subset\mC$ satisfy $\prod_{j=1}^{|\mC_{l_p}|}A_{l_p}^j=\alpha_{l_p}I$ for $p=1,2,...,k$, then the corresponding predictions $v(A_{l_p}^1)\cdot v(A_{l_p}^2)\cdot ...\cdot v(A_{l_p}^{|\mC_{l_p}|})=\alpha_{l_p}$ ($p=1,2,...,k$) hold both for NCHVTs and QM, and even do not depend on any particular preparation of the state. If we define the {\it reduced Bell operator} as $\mB^R\equiv\sum_{i\in\{1,...,s\}\backslash\{l_1,...,l_k\}}\alpha_i A_i^1\cdot A_i^2\cdot ...\cdot A_i^{|\mC_i|}$, then it is enough to test the following noncontextual inequality
\begin{align}\label{Noncontextual-ineq2}
    \mM^S(\mB^R):~\langle \mB^R\rangle\leq \beta_c(\mC)-\sum_{p=1}^k\alpha_{l_p}.
\end{align}
It can be considered as a special kind of symmetry.

Note that each term in the noncontextual inequality is just a correlation of all the observables in some context $\mC_i$,
and the test of quantum violation for each noncontextual inequality can be accomplished by the sequential measurements.
First, a {\it product partition} of $m$ compatible observables $(O^1, O^2,..., O^m)$ can be defined as
$(\prod_{i_1\in\mI^1}O^{i_1},\prod_{i_2\in\mI^2}O^{i_2},... ,\prod_{i_q\in\mI^r}O^{i_r})$ wherein $\cup_{i=1}^r\mI^i=\{1,2,...,m\}$ and
$\mI^i\cap\mI^j=\emptyset~(\forall i\neq j)$.
In a cascade measurement, the correlation $\langle A_i^1\cdot A_i^2\cdot ...\cdot A_i^{|\mC_i|}\rangle_{\mathrm{QM}}$ can be considered as the coincidence of any product partition of $|\mC_i|$ observables $(A_i^1, A_i^2,..., A_i^{|\mC_i|})$. This thought comes from the fact that in
a  multi-stage  measurement, one can combine some steps as a whole (e.g. stages (1,(23),4) resulting from stages (1,2,3,4) after some combination) for consideration. This gives us another kind of symmetry.
Based on this observation, we can define a useful cascade measurement equivalent relation.
Denote by a single-qubit Pauli operator $B_i~(i=1,2,...,n)$  for the $i$-th qubit, i.e., $B_i\in\{X_i,Y_i,Z_i\}$, then
a PO  $A_i$ is of the form  $A_i=(\otimes_{k\in\mI}B_{k})$, where $\mI$ are the index set for some qubits.
Two correlations of compatible POs,  $E=\langle (\otimes_{i_1\in\mI_1}B_{i_1})\cdot(\otimes_{i_2\in\mI_2}B_{i_2})\cdot...\cdot(\otimes_{i_p\in\mI_p}B_{i_p})\rangle$ and  $F=\langle (\otimes_{j_1\in\mJ_1}B_{j_1})\cdot(\otimes_{j_2\in\mJ_2}B_{j_2})\cdot...\cdot(\otimes_{j_q\in\mJ_q}B_{i_q})\rangle$, are {\it cascade  measurement equivalent}, if (i) the total numbers of any $B_i$ in the expression of $E$ and  $F$ are the same; (ii) for the POs  with respect to any pair of $(\mI_i, \mI_j)$ sharing two or more qubit indices in common,  the same indices should be
presented in the POs with respect to some $(\mJ_{i^{\prime}},\mJ_{j^{\prime}})$.

For example, the number of any Pauli operator $B_i$ in $\langle Y_1\cdot (X_2Y_3)\cdot (Y_2X_3)\cdot X_4\rangle$ and $\langle(Y_1X_2Y_3)\cdot(Y_2X_3X_4)\rangle$ are the same. Besides, the qubit indices $2$ and $3$ are shared by $(\mI_2,\mI_3)$ and $(\mJ_1,\mJ_2)$, where $\mI_2=\mI_3=\{2,3\}$ and $\mJ_1=\{1,2,3\},\mJ_2=\{2,3,4\}$. Therefore, $B_i$ in $\langle Y_1\cdot (X_2Y_3)\cdot (Y_2X_3)\cdot X_4\rangle$ and $\langle(Y_1X_2Y_3)\cdot(Y_2X_3X_4)\rangle$ are cascade  measurement equivalent.

For two GHZ-Peres-Mermin type proofs of KS theorem $\{\mC, |\varphi\rangle; \mM(\mB)\}$ and  $\{\mC^{\prime}, |\varphi\rangle; \mM^{\prime}(\mB^{\prime})\}$, if the expectation of each term in the reduced  Bell operators $\mB^R$ has its cascade measurement equivalent counterpart in ${\mB^R}^{\prime}$ and vice versa, their noncontextual inequalities $\mM(\mB)$ and  $\mM^{\prime}(\mB^{\prime})$ can be defined as {\it equivalent noncontextual inequalities}.
Accordingly, the associated state-dependent noncontextual models can be defined as  {\it equivalent proofs of KS theorem}.


These equivalent noncontextual inequalities, say $\{\mM(\mB), \mM^{\prime}(\mB^{\prime}), \mM^{\prime\prime}(\mB^{\prime\prime}), ...\}$, can be alternatively tested by the
quantum expectation of  their respective reduced Bell operators $\mB^R,{\mB^R}^{\prime},{\mB^R}^{\prime\prime},\cdots$. In other words,
One only need to test the corresponding simplified noncontextual inequalities $\mM^S(\mB^R),{\mM^S}^{\prime}({\mB^R}^{\prime}),{\mM^S}^{\prime\prime}({\mB^R}^{\prime\prime}),\cdots$. Their cascade measurement equivalent
items can obviously be tested in the same experiment, which indicates that $\mM^S(\mB^R),{\mM^S}^{\prime}({\mB^R}^{\prime}),{\mM^S}^{\prime\prime}({\mB^R}^{\prime\prime}),\cdots$ can
 be tested in the same experiment. 

\section{Experimental data}\label{app2}
To calculate the fidelity $\bar{F}_{\mathrm{exp}}$,
it only needs to measure the four elements of $\rho_{\mathrm{exp}}$ since other elements are zero in theory.
Two off-diagonal elements can be measured with projective measurements (see Table~\ref{table1}), whose real and imaginary part of each element can be written as
\begin{subequations}
  \begin{align}
    &\Re(\left<{000}\right|\rho_{\mathrm{exp}}\left|{111}\right>)={\Big(}\left<X^{\mathrm{o}}X^{\mathrm{s}}X^{\mathrm{p}}\right>-\left<X^{\mathrm{o}}Y^{\mathrm{s}}Y^{\mathrm{p}}\right>\notag\\
    &-\left<Y^{\mathrm{o}}X^{\mathrm{s}}Y^{\mathrm{p}}\right>-\left<Y^{\mathrm{o}}Y^{\mathrm{s}}X^{\mathrm{p}}\right>{\Big)}/8,\\
    &\Im(\left<{000}\right|\rho_{\mathrm{exp}}\left|{111}\right>)={\Big(}\left<Y^{\mathrm{o}}Y^{\mathrm{s}}Y^{\mathrm{p}}\right>-\left<X^{\mathrm{o}}X^{\mathrm{s}}Y^{\mathrm{p}}\right>\notag\\
    &-\left<Y^{\mathrm{o}}X^{\mathrm{s}}X^{\mathrm{p}}\right>-\left<X^{\mathrm{o}}Y^{\mathrm{s}}X^{\mathrm{p}}\right>{\Big)}/8.
  \end{align}
\end{subequations}
Another two diagonal elements, meaning the probabilities of $\left|{000}\right>$ and $\left|{111}\right>$ ($\left<{000}\right|\rho_{\mathrm{exp}}\left|{000}\right>$ and $\left<{111}\right|\rho_{\mathrm{exp}}\left|{111}\right>$), have been listed in the Table \ref{table1}.
Besides, the results of Pauli measurement $Z^{\mathrm{o}}Z^{\mathrm{s}}Z^{\mathrm{p}}$ are also list in Table \ref{table1}.

\nocite{*}
%

\end{document}